\documentclass[aip,pof,preprint]{revtex4-1}
\usepackage{epsfig,graphics,amssymb,amsmath,subeqnarray}
\usepackage{color}
\usepackage[utf8]{inputenc}

\def\F{{\mathbf{F}}}\def\u{{\mathbf{u}}}
\def\U{{\mathbf{U}}}\def\T{{\mathbf{T}}}
\def\I{{\mathbf{I}}}
\def\v{{\mathbf{v}}}\def\n{{\mathbf{n}}}\def\e{{\hat{\mathbf{e}}}}
\def\A{{\mathbf{A}}}\def\V{{\mathbf{V}}}
\def\r{{\mathbf{r}}}\def\sigmaB{{\boldsymbol{\sigma}}}
\def\OmegaB{{\boldsymbol{\Omega}}}\def\tauB{{\boldsymbol{\tau}}}
\def\nablaB{{\boldsymbol{\nabla}}}\def\gammaB{{\dot{\boldsymbol{\gamma}}}}
\def\gammaBs{{\dot{\boldsymbol{\gamma}^{*}}}}\def\varepsilon{\epsilon}
\def\De{{\text{De}}}\def\Deso{{\text{De}_{\text{so}}}}

\def\A{\stackrel{\triangledown}{\mathbf{A}}}
\def\Utaup{\stackrel{\triangledown}{\boldsymbol{\tau}^p}}
\def\Utaups{\stackrel{\triangledown}{\boldsymbol{\tau}^{p*}}}
\def\Utau{\stackrel{\triangledown}{\boldsymbol{\tau}}}
\def\Ugamma{{\stackrel{\triangledown}{\dot{\boldsymbol{\gamma}}}}}

\begin{document}

\title{Micropropulsion and microrheology in complex fluids via
symmetry breaking}

\begin{abstract}

Many biological fluids have polymeric microstructures and display non-Newtonian  rheology. We take advantage of such nonlinear fluid behavior and combine it with geometrical symmetry-breaking to design a novel small-scale propeller able to move only in complex fluids. Its propulsion characteristics are explored numerically in an Oldroyd-B fluid for finite Deborah numbers while the small Deborah number limit is investigated analytically using a second-order fluid model.  We then derive expressions relating the propulsion speed to the rheological properties of the complex fluid, allowing thus to infer  the normal stress coefficients in the fluid from the  locomotion of the propeller. Our simple mechanism can therefore be used either  as  a non-Newtonian micro-propeller or as a micro-rheometer.

\end{abstract}

\author{On Shun Pak}
\affiliation{
Department of Mechanical and Aerospace Engineering, 
University of California San Diego,
9500 Gilman Drive, La Jolla CA 92093-0411, USA.}

\author{LaiLai Zhu}
\affiliation{
Linn\'{e} Flow Center, KTH Mechanics, S-100 44 Stockholm, Sweden.
}

\author{Luca Brandt}
\email{luca@mech.kth.se}
\affiliation{
Linn\'{e} Flow Center, KTH Mechanics, S-100 44 Stockholm, Sweden.
}

\author{Eric Lauga}
\email{elauga@ucsd.edu}

\affiliation{
Department of Mechanical and Aerospace Engineering, 
University of California San Diego,
9500 Gilman Drive, La Jolla CA 92093-0411, USA.}

\date{\today}

\maketitle

\section{Introduction}\label{sec:introduction}

Life at low Reynolds numbers has attracted considerable attention in the past  few decades \cite{purcell77, brennen, fauci3, lauga2}. The absence of inertia plays a remarkable role in the swimming of microorganisms. In a Newtonian flow, the scallop theorem \cite{purcell77} constrains the types of locomotion strategies which are effective  in the microscopic world and reciprocal motion -- as are called those with  a time-reversal symmetry -- cannot lead to any net propulsion (or fluid transport).   Microorganisms evolved different propulsion strategies to achieve micro-propulsion, including the active propagation of flagellar waves for eukaryotic cells and the passive  rotation of rigid helical flagella for bacteria \cite{brennen}. 

The physics of low Reynolds number locomotion is relatively well explored in the Newtonian limit  (see reviews in Refs.~\cite{purcell77, brennen, fauci3, lauga2} and references therein). Beyond improving our understanding of biological processes, applications of these physical principles led to progress in the design of synthetic micro-swimmers for potential future biomedical applications  \cite{dreyfus, wang09, ebbens, lauga2,lauga_scallop}. In contrast, fundamental properties of life in complex, non-Newtonian, flows remain surprisingly unaddressed. Non-Newtonian flow behaviors can be appreciated through well-known manifestations from daily life, for example the climbing of dough up kitchen  mixing blades  (termed rod-climbing, or Weissenberg, effect) or the remarkable behavior of Silly Putty, a popular toy which bounces like a solid rubber ball when thrown to the floor but melts like a fluid when left on a surface for some time \cite{bird1, larson, morrison}. 

Many situations exist wherein microorganisms encounter biological fluids which have  polymeric microstructures and  non-Newtonian rheological properties. For example, spermatozoa swim through the viscoelastic cervical mucus and along the mucus-covered fallopian tubes \cite{katz78, katz80, katz81,suarez92,suarez06, fauci3}; cilia lie in a layer of mucus along the human respiratory tract \cite{sleigh}; \textit{Helicobacter pylori}, a bacterium causing ulcer, locomotes through mucus lining of the stomach \cite{montecucco}; spirochetes moves through host tissue during infection \cite{wolgemuth}; in biofilms, bacteria are embedded in cross-linked polymer gels \cite{otoole, donlan, costerton,costerton95,weitz}.

Physically and mathematically, the presence of polymeric stresses in a complex fluid means that the usual  properties associated with the absence of inertia in the Newtonian limit cease to be valid, in particular  kinematic reversibility and the linearity of the flow equations. In return, non-Newtonian effects such as stress relaxation, normal stress differences, and shear-rate dependent viscosity manifest themselves \cite{bird1,bird2,larson,morrison} . 

Past theoretical and experimental studies have investigated  the waveforms and swimming paths of microorganisms in complex fluids \cite{katz78, katz80, katz81, ishijima, suarez92,fauci3, lauga2, fu08, harman}. An active discussion in the biomechanics community has recently focused on the simple question: does fluid elasticity enhance or deteriorate propulsion at the microscopic scale? Theoretical studies on infinite models \cite{lauga1,powers07,powers09} showed that, for fixed body-frame kinematics, the propulsion speed decreases in a viscoelastic fluid. Numerical studies on a finite swimmer  \cite{shelley10} demonstrated that the propulsion speed could be enhanced by the presence of polymeric stress for some prescribed kinematics. Experimental investigations suggested evidences for both \cite{arratia11,liu11}. It was also shown that reciprocal actuation on a fluid, unable to provide net locomotion or flow transport in the Newtonian case, can be rendered effective by  viscoelasticity \cite{norman, lauga_life, pak10}. The presence of polymeric stress has also interesting  consequences on the rate of flagellar synchronization \cite{elfring}.

Normal stress differences in a complex fluid are responsible for a number of important non-Newtonian effects  \cite{bird1,larson,morrison} including  the rob-climbing effect mentioned above impacting many applications such as mixing, and the swelling of polymer melts when extruded from dies in manufacturing processes posing constraints on the rate of extrusion. In a pure shear flow with an arbitrary Reynolds number and a shear rate $\dot{\gamma}$, assuming that the flow is in the $x-$direction and the velocity varies in the $y-$direction, the $z-$direction being called the neutral direction \cite{bird1},  the first and second normal stress coefficients are defined as $\Psi_1  = (\tau_{xx} - \tau_{yy})/\dot{\gamma}^2$ and $\Psi_2  = (\tau_{yy} - \tau_{zz})/\dot{\gamma}^2$ respectively, where $\tau_{ij}$ are the components of the deviatoric stress tensor. In a Newtonian flow, there are no normal stress differences ($\Psi_1 = \Psi_2 = 0$), whereas for polymeric fluids typically $\Psi_1 > 0$ and $\Psi_2 < 0$. The magnitude of the second normal stress coefficient is usually much smaller than that of the first normal stress coefficient ($|\Psi_2| \ll \Psi_1$). In the rob-climbing phenomenon, both first and second normal stress coefficients contribute to the effect \cite{bird1}. However, due to its small magnitude, the effect of the second normal stress coefficient is shadowed by that of the first normal stress difference \cite{bird1}. The existence of the second normal stress difference can be demonstrated in a free-surface flow driven by gravity through a tilted trough: a Newtonian fluid has a flat free surface (with negligible meniscus effect), while the free surface of a non-Newtonian fluid becomes convex due to second normal stresses \cite{wineman, tanner, couturier}.

In this work, we propose a simple mechanism able to take advantage of the presence of normal stress differences  to propel in a complex fluid. Our geometry, shown in Fig.~\ref{fig:snowman}, consists of two linked small spheres propelling   under the action of an external torque, a setup we will  refer to as a ``snowman". Locomotion is enabled solely by the presence of normal stress differences, and no motion exists in a Newtonian environment, a fact that can in turn be used to infer the normal stress coefficients of a complex fluid. In essence, as complex fluids lead to new modes of small-scale propulsion, symmetrically the presence of propulsion in an environment can  be used to locally probe the rheological properties of the fluid.  

The first normal stress coefficient of a fluid can be measured directly from a conventional cone-and-plate rheometer \cite{bird1, larson, morrison}; the measurement of the second normal stress coefficient however has been a longstanding challenge \cite{brown, baek, kulkarni, schweizer}. A number of methods were proposed (see a review in Ref.~\cite{schweizer}), including a modified cone-and-plate rheometry with pressure transducers \cite{baek,bird1}, a subtle evaluation of a combination of cone-and-plate and parallel-plate experiments \cite{brown}, rheo-optical measurements \cite{brown, kulkarni}, and the use of a cone-and-partitioned plate tool \cite{schweizer}. Recently,   a microrheological technique was proposed to measure the first and second normal stress coefficients \cite{khair}. In microrheology, colloidal probes are either actively driven, or passively diffusing, and their dynamics allows to infer  local  rheological information.  Microrheology enjoys many advantages over conventional macroscopic rheological measurements \cite{mason,waigh}, including the reduction in sample size, the ability to probe spatially-inhomogeneous environments, and the possibility of performing measurements in living cells \cite{mason,waigh,weihs, wirtz}.  The mechanism we propose in this paper would be classified as ``active'' microrheology, a situation where colloidal probes are actively manipulated to drive the material out of equilibrium and probe its nonlinear mechanical properties  \cite{mason,squires08}. We offer in this paper an alternative microrheological technique capable of probing both first and second stress coefficients by using only kinematic measurements.

This paper is organized as follows. In Sec.~\ref{sec:setup}, we introduce the geometric and kinematic setup of our proposed mechanism and the polymeric fluid models adopted in our study. We first investigate in Sec.~\ref{sec:propulsion} the propulsion characteristics of the snowman in a complex fluid, followed in Sec.~\ref{sec:rheology} by the method of inferring the normal stress coefficients from its locomotion. We then provide a qualitative, and intuitive, explanation of the locomotion enabled by normal stresses in Sec.~\ref{sec:physical} before concluding the paper in Sec.~\ref{sec:discussion}.

\section{Setup}\label{sec:setup}
\subsection{Kinematics} \label{sec:kinematics}

\begin{figure}[t]
\begin{center}
\includegraphics[width=0.20\textwidth]{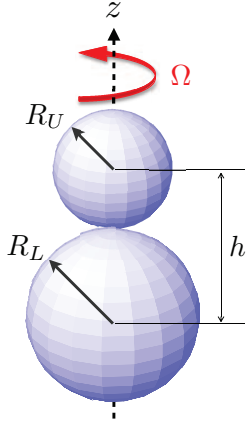}
\end{center}
\caption{\label{fig:snowman} Geometrical setup of two spheres (``snowman'') rotating with angular velocity $\Omega$ along their separation axis. The radii of the upper and lower spheres are denoted by $R_U$ and $R_L$ respectively. The centers of the spheres are separated by a distance, $h$ (for touching spheres, $h=R_U + R_L$). }
\end{figure}

By symmetry, the rotation of a single sphere in any homogeneous fluid produces no net locomotion. Inserting a second sphere, of different size, breaks the geometrical symmetry and can potentially allow locomotion. We first consider the rotation of two unequal spheres touching each other as a single rigid body (see the geometry and notations in Fig.~\ref{fig:snowman}), the ``snowman'' geometry. We label the line of centers of the spheres as the $z$-axis. Without loss of generality, we assume the radius of the upper sphere ($R_U$) is smaller than that of the lower sphere ($R_L \ge R_U$).  The distance between the centers of the spheres is denoted by $h$. For the case of touching spheres, we thus have $h = R_U+R_L$.

From a kinematic standpoint we assume that  the rigid body rotates with a steady angular velocity about the $z$-axis, $\boldsymbol{\Omega} = (0,0,\Omega >0)$, but is otherwise free to move. Given that the snowman is axisymmetric, the only direction it could potentially move is the $z$-direction. We  assume the net hydrodynamic force acting on the snowman is zero for all times (free-swimming condition), and aim at computing the rigid body (swimming) velocity necessary to maintain force-free motion.

In a Newtonian fluid without inertia, it is straightforward to  show using kinematic reversibility and reflection symmetry that a rotating snowman  cannot swim -- a result true for any degree of geometrical asymmetry. The central question at the heart of this paper is: Can elasticity of the fluid enable propulsion of the snowman? We answer this question in the following sections by studying the locomotion of a rotating snowman in polymeric fluids described by the two constitutive relations.

\subsection{Polymeric fluid dynamics}
We consider an incompressible low-Reynolds-number flow in a complex fluid. Denoting the velocity field as $\u$ and the
 fluid stress as $\sigmaB = -p \I + \tauB$, where $p$ is the pressure, and $\tauB$ is the deviatoric stress tensor, the
conservation of mass and momentum are given by the continuity equation and Cauchy's equation of motion respectively
\begin{align}
\nablaB \cdot \u &= 0, \\
\nablaB \cdot \sigmaB &= 0.
\end{align}
For closure, we require a constitutive equation relating the deviatoric stresses $\tauB$ to the kinematics of the flow.
Obviously a large number of models have been proposed in the past to describe polymeric fluids. In this work two constitutive equations are used  to study the viscoelastic locomotion of a snowman.

\subsubsection{Oldroyd-B fluid}
The classical Oldroyd-B constitutive equation is arguably the most famous constitutive model for polymeric fluids \cite{bird1,bird2,larson,morrison}. It has a sound physical origin and can be derived from a kinetic theory of polymers in the dilute limit by modeling polymeric molecules as linearly elastic dumbbells. The predictions also agree well with experimental measurements  up to order one Weissenberg numbers, although it is known to suffer deficiencies for larger values \cite{bird1, bird2, larson,morrison}. In an Oldroyd-B fluid, the deviatoric stress is the sum of  two components, $\tauB = \tauB^s + \tauB^p$, where $\tauB^s$ and $\tauB^p$ denote, respectively, the Newtonian solvent contribution and polymeric contribution to the stress. The constitutive relation for the Newtonian contribution is given by $\tauB^s = \eta_s \gammaB$, where  $\gammaB = \nablaB \u + \nablaB \u^T$ is the rate of strain tensor and $\eta_s$ is the solvent contribution to the viscosity.  {The momentum equation can thus be written as
\begin{align}
-\nablaB p + \eta_{s}  \nablaB \cdot \gammaB   + \nablaB \cdot \tauB^p = 0,
\end{align}
}
The polymeric stress $\tauB^p$ is then assumed to be governed by the upper-convected Maxwell equation
\begin{align}
\tauB^p + \lambda \Utaup = \eta_p \gammaB,  \label{eqn:polymeric}
\end{align}
where $\lambda$ is the polymeric relaxation time and $\eta_p$ is the polymer contribution to the viscosity  \cite{bird1,larson,morrison}. The upper-convected derivative for a tensor $\mathbf{A}$ is defined as
\begin{align}
\A = \frac{\partial \mathbf{A}}{\partial t} + \u \cdot \nablaB \mathbf{A} - (\nablaB \u^T \cdot \mathbf{A} + \mathbf{A} \cdot \nablaB \u),
\end{align}
which calculates the rate of change of $\mathbf{A}$ while translating and deforming with the fluid.

Combining the Newtonian and polymeric constitutive relations, we obtain the Oldroyd-B constitutive equation for the total stress, $\tauB$, as
\begin{align}
\tauB + \lambda \Utau = \eta (\gammaB + \lambda_2  \Ugamma), \label{eqn:OB}
\end{align}
where the total viscosity is given by $\eta = \eta_s + \eta_p$, and $\lambda_2 = \lambda \zeta$ denote the retardation times (we define the relative viscosity $\zeta = \eta_s/\eta < 1$). For steady shear  of an Oldroyd-B fluid, both the viscosity and the first normal stress coefficient are constant, and the second normal stress coefficient is zero \cite{bird1}. The Oldroyd-B fluid is the model we will use for our numerical approach.

\subsubsection{Second-order fluid} \label{sec:secondOrder}

For slow and slowly varying flows, the second-order fluid model applies. It is the first non-Newtonian term
in a systematic asymptotic expansion of the relationship between the stress and the rate of strain tensors called the
retarded-motion expansion. It describes small departures from Newtonian fluid behavior, and the instantaneous
constitutive equation is given in this model by
\begin{align}
\tauB = \eta \gammaB - \frac{1}{2} \Psi_1 \Ugamma + \Psi_2 (\gammaB \cdot \gammaB),
\end{align}
where $\Psi_1$ and $\Psi_2$ are the first and second normal stress coefficients respectively. Note that if $\lambda =
0$ while $\lambda_2\neq 0$ in the Oldroyd-B model, Eq.~\eqref{eqn:OB}, it reduces to a second-order fluid with a vanishing second normal stress coefficient ($\Psi_2=0$) \cite{bird1}. The second-order fluid model will enable us to derive theoretically the behavior of the snowman for small deformations.

\subsection{Non-dimensionalization}

We non-dimensionalize lengths by the radius of the lower sphere $R_L$, times by $1/\Omega$, and use the total fluid viscosity, $\eta$, to provide the third fundamental unit. Hence, velocities, shear rates, and stresses are scaled by $R_L \Omega$, $\Omega$, and $\eta \Omega$ respectively. The dimensionless radius of the upper  sphere becomes then $r^* = R_U/R_L$ while the lower sphere has now radius 1. We have $h^* = h/R_L$ denoting the dimensionless distance between the centers of the sphere ($h^* = 1+ r^*$ for two touching spheres). Both spheres rotate at the same dimensionless unit speed, $\Omega^* =1$. The starred variables represent dimensionless variables in this paper. The Deborah number \cite{bird1, larson, morrison}, $\De = \lambda \Omega$, is a dimensionless number defined as the ratio of a characteristic time scale of the fluid (the polymeric relaxation time, $\lambda$) to a characteristic time scale of the flow system ($1/\Omega$), and appears  in the dimensionless momentum equation and upper-convected Maxwell equation 
\begin{align}
-\nablaB p^{*} + \zeta \nablaB \cdot \gammaB^{*}   + \nablaB \cdot \tauB^{p*} = 0\label{eqn:MomDimensionless},\\
\tauB^{p*} + \De \Utaups = \left(1-\zeta \right) \gammaBs.  \label{eqn:UCMDimensionless}
\end{align}
The limit $\De =0$ corresponds to a Newtonian fluid.

Alternatively, the upper-convected Maxwell equation of the polymeric stress, Eq.~\eqref{eqn:UCMDimensionless}, can be combined with the constitutive relation of the Newtonian contribution to obtain the dimensionless Oldroyd-B constitutive equation for the total stress $\tauB^*$ as
\begin{align}
\tauB^* + \De \stackrel{\triangledown}{\boldsymbol{\tau}^*} = \gammaB^* + \De_2 \stackrel{\triangledown}{\dot{\boldsymbol{\gamma}}^*}, \label{eqn:OBDimensionless}
\end{align}
where we have defined another Deborah number, $\De_2$, in terms of the retardation time, $\De_2 = \lambda_2 \Omega = \De \zeta$.

The dimensionless constitutive relation for a second-order fluid is now given by
\begin{align}
\tauB^* = \gammaB^* - \De_{\text{so}} \left( \stackrel{\triangledown}{\dot{\boldsymbol{\gamma}}^*} + B \gammaB^* \cdot \gammaB^* \right), \label{eqn:secondOrderDimensionless} 
\end{align}
where we have defined another Deborah number for the second-order fluid, namely $\De_{\text{so}} = \Psi_1 \Omega/2 \eta$, and $B = -2 \Psi_2/ \Psi_1 \ge 0$.

Importantly, we note that the definition of the Deborah number of an Oldroyd-B fluid is different from that of a second-order fluid, because the relaxation time of an Oldroyd-B fluid is defined only by the polymer, whereas the relaxation time of a second-order is defined by both the polymer and the solvent \cite{lee10}. The two Deborah numbers are related by the relation $\De_{\text{so}} = \De(1-\zeta)$. We shall mostly use the Deborah number defined for an Oldroyd-B fluid ($\De$) for the presentation of our final results, since we feel it is the one with  the most intuitive definition. The Oldroyd-B equation is valid up to moderate $\De$, and the second order fluid is valid for small $\De$ (or $\De_{\text{so}}$), and we thus expect the results from both models to match when $\De$ (or $\De_{\text{so}}$) is sufficiently small.

\section{Propulsion of snowman in a complex fluid}\label{sec:propulsion}

As argued in Sec.~\ref{sec:kinematics}, asymmetry alone does not lead to net locomotion upon rotating a snowman in a Newtonian fluid. We now explore the effects of fluid elasticity on the propulsion of a snowman: Does it  even move? Which direction does it go? And how fast? Using the Oldroyd-B fluid model, we first explore numerically the propulsion characteristics of the snowman from small to moderate Deborah numbers. Next, the small $\De$ limit is studied analytically via the second-order fluid model.

\subsection{Moderate Deborah number} \label{sec:moderateDe}

{We employed a finite element model to compute the polymeric flow as described by Eqs.~\eqref{eqn:MomDimensionless} and 
\eqref{eqn:UCMDimensionless}.  A formulation called the Discrete Elastic-Viscous Split Stress (DEVSS-G)~\cite{devssFortin,Devss-G98} is implemented here to improve numerical stability.}
{The momentum equation, Eq.~\eqref{eqn:MomDimensionless},  is rewritten as
\begin{align}\label{eqn:devssg}
 \nablaB \cdot \mu_{a}(\nablaB \u^{*} + \nablaB \u^{*T})-\nablaB p^{*} + \nablaB \cdot
\boldsymbol{\tau}^{p*}-\nablaB \cdot (\mu_{a} - \zeta)(\mathbf{G} + \mathbf{G}^{T})=0,
\end{align}
where the tensor $\mathbf{G} \equiv \nablaB \u^{*}$ is introduced as a finite element approximation of the velocity gradient tensor $\nablaB \u^{*}$. An additional elliptic term, $\nablaB \cdot \mu_{a}(\nablaB \u^{*} +\nablaB \u^{*T}) - \nablaB \cdot \mu_{a}(\mathbf{G} + \mathbf{G}^{T})$, is added into the momentum equation for stabilization \cite{MIT99_DAVSSDG}. In the limit that the mesh size in the finite element approximation tends to zero, $\mathbf{G}$ approaches $\nablaB \u^{*}$ and the elliptic term vanishes, reducing Eq.~\eqref{eqn:devssg} to Eq.~\eqref{eqn:MomDimensionless}.
$\mathbf{G}$ is also used to approximate the velocity gradient term  $\nablaB \u^{*}$ in the constitutive equation, Eq.~\eqref{eqn:UCMDimensionless}. For simulations in this work, we choose $\mu_{a}=1$ as in Liu \textit{et al.}~\cite{Devss-G98}.

A Galerkin method is used to discretize the momentum equations, continuity equation, and the
equation for the additional unknown $\mathbf{G}$.  Quadratic elements are used for $\u^{*}$ and linear elements for both
$p^{*}$ and $\mathbf{G}$. The streamline-upwind/Petrov-Galerkin(SUPG) \cite{supgCrochet} method is adopted to
discretize the constitutive equation, Eq.~\eqref{eqn:UCMDimensionless}, to improve numerical stability. The resulting weak form of the model  is formulated as
\begin{align}\label{eqn:weakpoly}
\{\mathbf{S}+\frac{h_c}{U_c}\u^{*} \cdot \nabla \mathbf{S},\,\, \tauB^{p*} & + \De (\u^{*} \cdot \nabla \tauB^{p*}
- \mathbf{G}^{T} \cdot \tauB^{p*} - \tauB^{p*} \cdot \mathbf{G} )
-\ (1-\zeta\ )(\mathbf{G}+\mathbf{G}^{T} ) \} =0,
\end{align}
where $\mathbf{S}$ denotes the test function for $\tauB^{p*}$, $h_c$ is a characteristic mesh size, and
$U_c$ is the magnitude of a local characteristic velocity (we choose the norm of $\u^{*}$ as $U_c$). The framework for the implementation is provided by the commercial software COMSOL, which was successfully used for simulating the locomotion of squirmers in a viscoelastic fluid at low Reynolds numbers \cite{laipof1}.

We perform three-dimensional axisymmetric simulations on a two-dimensional mesh constructed with triangle elements.
Sufficiently refined mesh is generated near rotating objects to resolve the thin stress boundary layers, 
necessary to overcome numerical instabilities~\cite{viscoeReview,computaRheologyCFD} and improve accuracy.
We  validate our implementation by comparing numerical and analytical values of the hydrodynamic torque
on a rotating sphere in the Newtonian fluid. For the viscoelastic model, we validate our approach against
the simulations in Lunsmann \textit{et al.} \cite{Lunsmann1993} of a sedimenting sphere in a tube filled with Oldroyd-B fluid and the
analytical results in Bird \textit{et al.} \cite{bird1} of a rotating sphere in a second-order fluid.

Equipped with our computational model, we are able to show that  fluid elasticity does indeed enable the propulsion of the snowman provided the two spheres  have unequal sizes ($r^* < 1$). The snowman always swim in the positive $z$-direction (see Fig.~\ref{fig:snowman}), \textit{i.e.}~from the larger to  the smaller sphere. For illustration, we compute the dimensionless
 propulsion speed, $U^* = U/R_L\Omega$, of a typical snowman ($r^* = R_U/R_L = 0.5$) as a function of the Deborah number,
$\De$ (dot-dashed line - red online, Fig.~\ref{fig:moderateDe}), for a fixed relative viscosity $\zeta = 0.5$. 
When $\De =0$, the fluid reduces to the Newtonian limit and we recover that no propulsion is possible in this case. For small values of $\De$, the propulsion speed appears to grow linearly with $\De$, a result confirmed  analytically in the next section. A maximum swimming speed is reached at $\De \approx 1.75$, before decaying as $\De$ continues to increase. 

\begin{figure}[t]
\begin{center}
\includegraphics[width=1\textwidth]{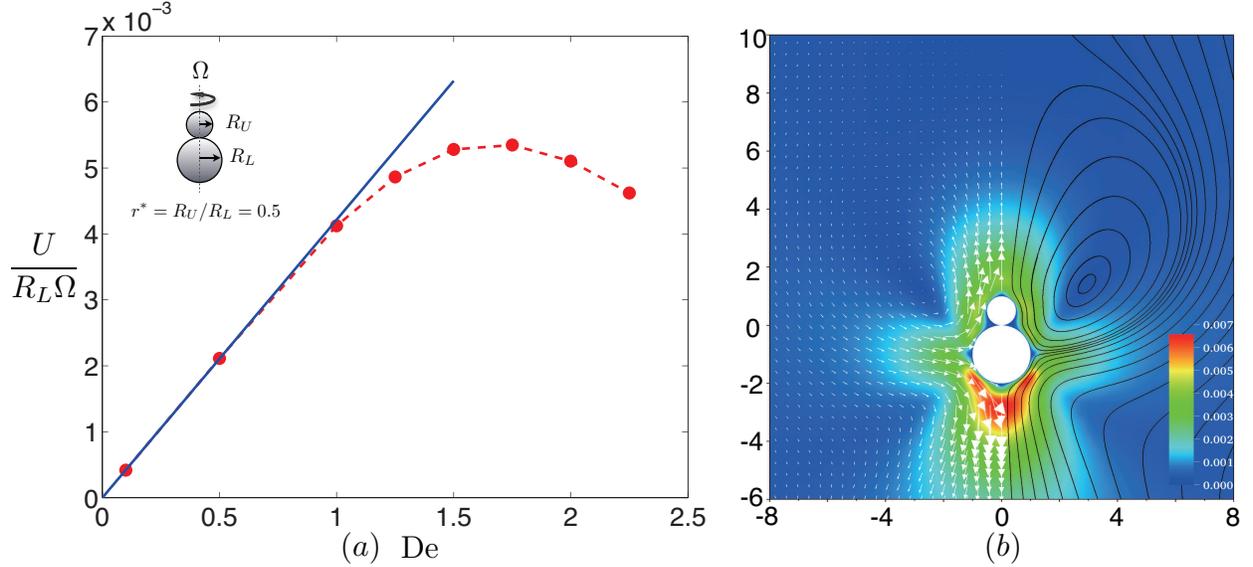}
\end{center}
\caption{\label{fig:moderateDe} Demonstration of snowman locomotion. In the case $r^* = R_U/R_L =0.5$ and $\zeta =0.5$, we plot: (a) Dimensionless propulsion speed, $U/R_L \Omega$, as a function of the Deborah number, $\De$. Dot-dashed line (red online): numerical simulations in an Oldroyd-B fluid; solid line (blue online): theoretical calculation using the reciprocal theorem in a second-order fluid, Eq.~\ref{eqn:finalRe}; (b) The streamline pattern and speed (shaded/color map) of the secondary flow for $\De = 0.1$ (streamline patterns at higher $\De$ are qualitatively similar).}
\end{figure}

In addition to the primary flow (the Newtonian component, $\De=0$), elastic stresses  around the snowman generate a secondary flow, understood simply as  the difference between the total flow and the Newtonian component. A typical secondary flow pattern  is shown {in the frame of the snowman} in Fig.~\ref{fig:moderateDe}b {($\De = 0.1$ and $\zeta=0.5$)}. {We depict the velocity vectors and streamlines with the shaded/color map representing the flow speed. Fluid is drawn towards the snowman parallel to the equatorial plane and then expelled along the axis, while a ring vortex is detected in the front. The maximum speed of the secondary flow is observed at the rear of the snowman, only about $0.7\%$ of the characteristic speed of the primary flow $R_{L} \Omega$. }

\subsection{Small Deborah number}\label{2nd_order}
To provide a theoretical approach to the snowman locomotion and to quantify the  connection between locomotion and rheology in the following sections we now consider the second-order fluid, which we remind is valid in the small-$\De$  limit  only  (Eq.~\ref{eqn:secondOrderDimensionless}). All variables are expanded in powers of the Deborah number,  $\Deso$, as
\begin{align}
\sigmaB &= \sigmaB_0 + \Deso \sigmaB_1 + O(\Deso^2), \\
\u &= \u_0 + \Deso \u_1 + O(\Deso^2),\\
\gammaB &= \gammaB_0 + \Deso \gammaB_1 + O(\Deso^2),\\
\U &= \U_0 + \Deso \U_1 + O(\Deso^2), 
\end{align}
where $\U$ denotes the propulsion velocity, $\U = (0,0,U)$. Other variables are expanded similarly. We drop the stars hereafter for simplicity, and all variables in this section are dimensionless unless otherwise stated. The locomotion problem is then solved order by order.

\subsubsection{Zeroth-order solution}

The zeroth order solution, \{$\sigmaB_0 = -p_0 \I + \gammaB_0, \u_0$\}, satisfies the Stokes equations, 
\begin{align}
\nablaB \cdot \sigmaB_0 &= \mathbf{0}, \\
\nablaB \cdot \u_0 &= 0,
\end{align}
where $\sigmaB_0 = -p_0 \I + \gammaB_0$. This is the Newtonian flow for two touching spheres rotating (at a rate of $\OmegaB$) about the line of their centers ($z$-axis). The exact solution in terms of analytical functions was  given by Takagi \cite{takagi74}  in tangent-sphere coordinates. No propulsion occurs in the Newtonian limit, $\U_0 = \mathbf{0}$, as expected. 

\subsubsection{First-order solution} \label{sec:FirstOrder}

The first order solution ($\sigmaB_1, \u_1$) of the main problem satisfies 
\begin{align}
\nablaB \cdot \sigmaB_1 &= \textbf{0} \label{eqn:firstOrder}, \\
\nablaB \cdot \u_1 &= 0,
\end{align}
where 
\begin{align}
\sigmaB_1 = - p_1 \I + \gammaB_1 -  \Ugamma_0 - B \gammaB_0 \cdot \gammaB_0. \label{eqn:firstSecond}
\end{align}
To compute the value of the first order propulsion velocity, $\U_1$, we will use a version of the reciprocal theorem for Stokes flows adapted to self-propulsion in viscoelastic fluids \cite{ho, chan, brunn76a, brunn76b, leal75, leal80, philips96, lauga_life, khair}.

Consider an auxiliary problem with identical geometry,  $\{\sigmaB_{\text{aux}}, \u_{\text{aux}}\}$, satisfying
\begin{align}
\nablaB \cdot \sigmaB_{\text{aux}} &= \mathbf{0}, \label{eqn:aux} \\ 
\nablaB \cdot \u_{\text{aux}} &= 0.
\end{align}
Taking the inner product of Eq.~\eqref{eqn:firstOrder} with $\u_{\text{aux}}$, minus the inner product of Eq.~\eqref{eqn:aux} with $\u_{1}$, and integrating over the entire fluid volume, we have trivially
\begin{align}
\int_{V_f} \u_{\text{aux}} \cdot (\nablaB \cdot \sigmaB_1) - \u_{1} \cdot (\nablaB \cdot \sigmaB_{\text{aux}}) dV = 0.
\end{align}
Using vector calculus we can rewrite the integral in the following form
\cite{leal07}
\begin{align}
\int_{V_f}  \nablaB \cdot (\u_{\text{aux}} \cdot \sigmaB_{1} - \u_{1} \cdot \sigmaB_{\text{aux}})dV =\int_{V_f} ( \nablaB \u_{\text{aux}}: \sigmaB_{1}-\nablaB \u_{1} : \sigmaB_{\text{aux}})  dV. \label{eqn:integral}
\end{align}
The left-hand side of Eq.~\eqref{eqn:integral}  can be converted to a sum of surface integrals by the divergence theorem while
the right-hand side can be simplified using the first-order constitutive equation, Eq.~\eqref{eqn:firstSecond}, leading to
\begin{align}\label{integral2}
\sum_{\alpha} \int_{S_{\alpha}} \n \cdot (\u_{\text{aux}} \cdot \sigmaB_{1}-\u_{1} \cdot \sigmaB_{\text{aux}}) dS = \int_{V_f} \left[ \left( \Ugamma_{0} + B \gammaB_0 \cdot \gammaB_0 \right) : \nablaB \u_{\text{aux}} \right] dV,
\end{align}
where $S_\alpha$ denotes the surface of different spheres ($\alpha=1,2$) and $\mathbf{n}$ represents the outward normal vector on the surface. The important simplification which took place in the right hand-side of Eq.~\eqref{eqn:integral} is that all Newtonian terms included in $\sigmaB_{\text{aux}}$ and $\sigmaB_{1}$ have canceled each other out by symmetry, and thus the only piece remaining in the right-hand side of Eq.~\eqref{integral2} is the non-Newtonian contribution  \cite{chan, philips96}.

Now, let $\U_1$ and $\OmegaB_1$ be the (unknown) first order translational and rotational velocities of the spheres in our main problem, while the translational and rotational velocities of the spheres in the auxiliary problem (known) are given by $\U_{\text{aux}}$ and $\OmegaB_{\text{aux}}$. On the surface $S_{\alpha}$ of one sphere, the no-slip and no-penetration boundary conditions lead to
\begin{align}
\u_{\text{aux}} &= \U_{\text{aux}} + \OmegaB_{\text{aux}} \times \r,\\
\u_{1} &= \U_{1} + \OmegaB_{1} \times \r,
\end{align}
where $\r$ is the position vector describing the surface. The integral relation, Eq.~\eqref{integral2},  becomes
\begin{align}\label{integral3}
\notag\sum_{\alpha} \U_{\text{aux}}^\alpha \cdot \int_{S_{\alpha}} \n \cdot \sigmaB_1 dS + \OmegaB_{\text{aux}}^\alpha \cdot \int_{S_{\alpha}} \r \times (\n \cdot \sigmaB_1)dS\\
 -\U_{1}^\alpha \cdot \int_{S_{\alpha}} \n \cdot \sigmaB_{\text{aux}} dS - \OmegaB_{1}^\alpha \cdot \int_{S_{\alpha}} \r \times (\n \cdot \sigmaB_{\text{aux}}) dS \notag \\
=  \int_{V_f} \left[ \left( \Ugamma_{0} + B \gammaB_0 \cdot \gammaB_0 \right) : \nablaB \u_{\text{aux}} \right] dV.
\end{align}

In Eq.~\eqref{integral3}, the integrals $ \int_{S_{\alpha}} \n \cdot \sigmaB_1 dS$ and $\int_{S_{\alpha}} \r \times (\n \cdot \sigmaB_1) dS$ represent the net hydrodynamic force and torque acting on the sphere $\alpha$ by the first order flow field. Let us denote $\F_1^{\alpha} = -\int_{S_{\alpha}} \n \cdot \sigmaB_1 dS$ and $\T_1^{\alpha}= - \int_{S_{\alpha}} \r \times (\n \cdot \sigmaB_1) dS$ 
 the net external force ($\F_1^{\alpha}$)  and external torque ($\T_1^{\alpha}$) acting on each sphere; the appearance of a minus sign comes from the fact that the total force and torque (external + fluid) acting on a body  have to sum to zero in the absence of inertia. In the free-swimming case there is an additional stronger constraint, namely  the total external force (or equivalently, the total fluid force) has to remain zero at all instant (we will enforce this constraint shortly). Defining also $\F_{\text{aux}}^{\alpha}$ and $\T_{\text{aux}}^{\alpha}$ as the external force and torque required to balance the fluid drag and torque on each sphere in the auxiliary problem we see that Eq.~\eqref{integral3} is transformed into
\begin{align}
 \sum_{\alpha} -\U_{\text{aux}}^\alpha \cdot \F_1^{\alpha} - \OmegaB_{\text{aux}}^\alpha \cdot \T^{\alpha}_1  +\U_{1}^\alpha \cdot \F_{\text{aux}}^{\alpha} + \OmegaB_{1}^\alpha \cdot \T_{\text{aux}}^{\alpha} =  \int_{V_f} \left[ \left( \Ugamma_{0} + B \gammaB_0 \cdot \gammaB_0 \right) : \nablaB \u_{\text{aux}} \right] dV. \label{eqn:generalRe}
\end{align}

The above relation remains actually true for any number of spheres and kinematics.  In the case of a snowman, we have two spheres  connected as a rigid body in both the main and auxiliary problems, hence $\U_1^1 = \U_1^2 = \U_1$, $\OmegaB^1=\OmegaB^2 = \OmegaB_1$, $\U_{\text{aux}}^1=\U_{\text{aux}}^2 = \U_{\text{aux}}$, and  $\OmegaB_{\text{aux}}^1=\OmegaB_{\text{aux}}^2 = \OmegaB_{\text{aux}}$. In the main problem we impose a rotational rate $\OmegaB$ on the snowman, which has been accounted for in the zeroth order (Newtonian) solution,  hence $\OmegaB_n^{\alpha} = 0$ for all $n \ge 1$. In addition, we define in the main problem the total external force and torque acting on the rigid body as $\F_1 = \F^1 + \F^2$, $\T_1 = \T^1+\T^2$, and in the auxiliary problem $\F_{\text{aux}} = \F_{\text{aux}}^1 + \F_{\text{aux}}^2$, and $\T_{\text{aux}} = \T_{\text{aux}}^1+\T_{\text{aux}}^2$. Using these simplifications the general relation, Eq.~\eqref{eqn:generalRe},  simplifies to
\begin{align}
- \left( \U_{\text{aux}} \cdot \F_1 + \OmegaB_{\text{aux}} \cdot \T_1  \right) + \U_1 \cdot \F_{\text{aux}}   =  \int_{V_f} \left[ \left( \Ugamma_{0} + B \gammaB_0 \cdot \gammaB_0 \right) : \nablaB \u_{\text{aux}} \right] dV \cdot
\end{align}

We now need to find  an auxiliary problem that facilitates the determination of the first order propulsion velocity, $\U_1$, in the main problem. An appropriate candidate is the translation of two touching spheres along the line of their centers without rotation, $\OmegaB_{\text{aux}} = 0$. The exact analytical solution was given by  Cooley and O'Neill \cite{cooley69}. By choosing this auxiliary problem, the relation further simplifies to
\begin{align}
- \U_{\text{aux}} \cdot \F_1 + \U_1 \cdot \F_{\text{aux}}   =  \int_{V_f} \left[ \left( \Ugamma_{0} + B \gammaB_0 \cdot \gammaB_0 \right) : \nablaB \u_{\text{aux}} \right] dV \cdot
\end{align}
If we do not allow the spheres to translate along the $z$-axis, $\U_1 =0$, an external force, $\F_1$, is required to hold the snowman in place given by 
\begin{align}
- \U_{\text{aux}} \cdot \F_1  =  \int_{V_f} \left[ \left( \Ugamma_{0} + B \gammaB_0 \cdot \gammaB_0 \right) : \nablaB \u_{\text{aux}} \right] dV \cdot
\end{align}

On the other hand, if we allow the snowman to translate freely without imposing any external forces, $\F_1 = 0$, then the first order propulsion velocity, $\U_1$, can be determined from
\begin{align}
\U_1 \cdot \F_{\text{aux}}   =  \int_{V_f} \left[ \left( \Ugamma_{0} + B \gammaB_0 \cdot \gammaB_0 \right) : \nablaB \u_{\text{aux}} \right] dV, 
\end{align}
where both $\F_{\text{aux}}$ and the integral are expressed in terms of known Newtonian solutions of the main and auxiliary problems. Since the propulsion velocity $\U_1$ (with magnitude $U_1$) and the force in the auxiliary problem $\F_{\text{aux}}$ (with magnitude $F_{\text{aux}}$) act both vertically, the first order propulsion speed is finally given by
\begin{align}
U_1 = \frac{1}{F_{\text{aux}}} \int_{V_f} \left[ \left( \Ugamma_{0} + B \gammaB_0 \cdot \gammaB_0 \right) : \nablaB \u_{\text{aux}} \right] dV, \label{eqn:finalRe}
\end{align}
where a positive value represents upward propulsion. 

Using Eq.~\eqref{eqn:finalRe} with the zeroth-order solution  \cite{takagi74}  and the auxiliary Newtonain solution \cite{cooley69} we are able to determine theoretically the  leading order propulsion speed of the snowman, $U = \Deso U_1 + O(\Deso^2) = \De (1-\zeta) U_1  + O(\De^2) $. The quadrature is performed in the tangent-sphere coordinates, with somewhat lengthy differential operations in evaluating the integrand. Our asymptotic results are shown in Fig.~\ref{fig:moderateDe} as a solid line (blue online). We see that our results 
predict very well the  propulsion speed of the snowman for small $\De $ when compared with numerical computations of the  Oldroyd-B fluid (dot-dashed line - red online, in  Fig.~\ref{fig:moderateDe}), and the agreement is excellent up to $\De \sim 1$. 

Note that in order to compare the results between the second-order fluid calculation and the Oldroyd-B numerics, the dimensionless parameter $B = - 2\Psi_2/\Psi_1$ in the second-order fluid has to be  taken to be zero because the  second normals stress coefficient is zero in the Oldroyd-B model.  
Experimentally, indeed we have $B \ll 1$.  
Mathematically, the propulsion velocity varies linearly with $B$, and a transition of propulsion direction occurs at $B = 1$. Such a transition also occurs in the direction of radial flow for  a single rotating sphere in a second-order fluid (see Sec.~\ref{sec:physical} for a related discussion).

\subsection{Propulsion characteristics}

Anticipating the section where we make the link between rheology and locomotion, we now investigate the impact of the snowman geometry on its propulsion performance in the low-$\De$ regime where our asymptotic results via reciprocal theorem are quantitatively accurate. 

\subsubsection{Touching spheres}

In the case of two touching sphere ($h^* = 1+ r^*$), the only free dimensionless geometric parameter is the ratio of the radius of the upper to that of the lower spheres $r^* = R_U/R_L \in [0,1]$. In the limit $r^*=0$, the snowman reduces to a single sphere, while the limit $r^*=1$ corresponds to two equal touching spheres; in both cases, there is no propulsion by symmetry. We therefore expect an optimal ratio $r^*$ for a maximum propulsion speed.  Using the reciprocal theorem, Eq.~\eqref{eqn:finalRe}, we calculate the propulsion speed as a function of $r^*$ (Fig~\ref{fig:touchingR}, solid line - blue online) and compare with the numerical results in an Oldroyd-B fluid (Fig.~\ref{fig:touchingR}, dots - red online) at $\De = 0.1$ and a typical relative viscosity $\zeta = 0.5$. The asymptotic results agree very well with the Oldroyd-B computations. The optimal sphere size ratio occurs at $r^*_{\text{opt}} \approx 0.58$. In addition to our computations and theoretical calculations, and based on physical understanding of the behavior of a single rotating sphere in a second-order fluid, a simplified analytical model can be constructed to predict the snowman dynamics with results shown as a dotted line (black online) in Fig.~\ref{fig:touchingR}; the details of this simple model are given in  Sec.~\ref{sec:physical}.

\begin{figure}[t]
\begin{center}
\includegraphics[width=0.45\textwidth]{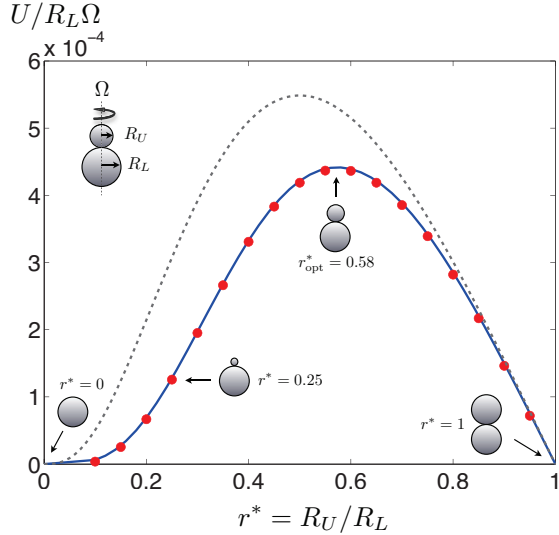}
\end{center}
\caption{\label{fig:touchingR} Propulsion speed of snowman with two touching spheres as a function of the ratio of radii at $\De=0.1$ and $\zeta = 0.5$. Dots (red online): numerical results in an Odroyd-B fluid. Solid line (blue online): theoretical calculation for  second-order fluid. Dotted line (black online):  simplified analytical model (Eq.~\ref{eqn:simpleTouching}).}
\end{figure}

\subsubsection{Separated spheres} \label{sec:separatedSphere}

Next, we let the two spheres be separated at a distance $h^*> 1+ r^*$ (no longer touching). The two spheres still rotate at the same speed as a rigid body and the separation distance is kept fixed by connecting the spheres with a drag-less slender rigid rod (a mathematically phantom rod) with negligible hydrodynamic contribution. Experimentally, this may be realized using, for example, using  nanowires \cite{fnote}. To compute the propulsion speed by the method described in Sec.~\ref{sec:FirstOrder} and therefore  Eq.~\eqref{eqn:finalRe}, we need two new Newtonian solutions, namely the zeroth-order solution and the auxiliary problem. The zeroth order solution  considers two separated unequal spheres rotating at the same rate in a Newtonian fluid, the exact solution of which was given by Jeffery \cite{jeffery15}  in bi-spherical coordinates. The appropriate auxiliary problem is  the translation in a Newtonian fluid of the same two-sphere geometry   along their axis of symmetry. Stimson and Jeffery \cite{stimson26} calculated that exact solution also in bi-spherical coordinates.

\begin{figure}[t]
\begin{center}
\includegraphics[width=0.9\textwidth]{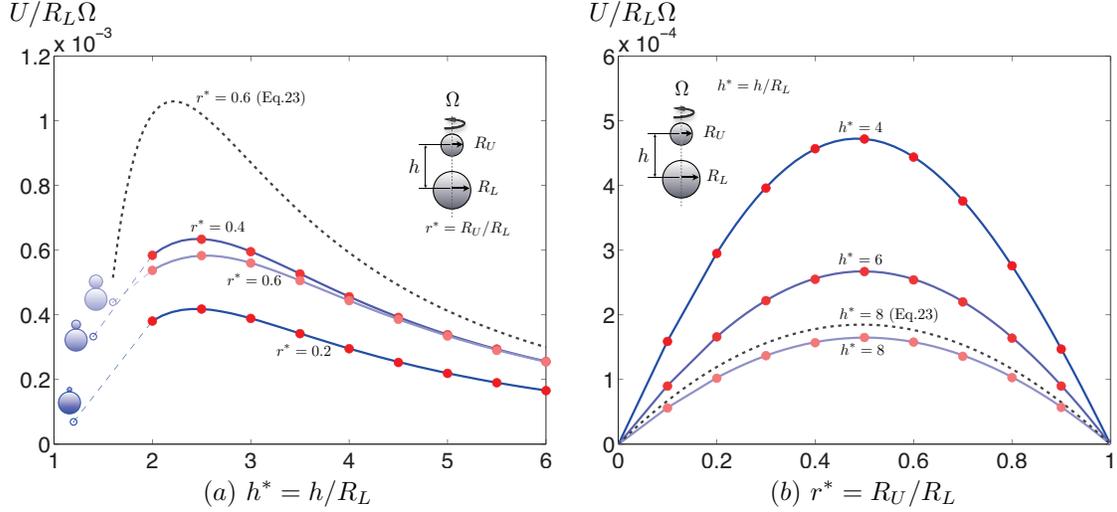}
\end{center}
\caption{\label{fig:separated} Propulsion speed of a separated snowman as a function of (a) the separation distance, and (b) ratio of the radii, at $\De=0.1$ and $\zeta = 0.5$. Dots (red online): numerical results in an Odroyd-B fluid. Solid line (blue online): second-order fluid analytical calculation. Dotted line (black online):  simplified model (Eq.~\ref{eqn:simple}).}
\end{figure} 

For very separated spheres $h^* \gg 1$, the propulsion is expected to decay with the separation distance. Hydrodynamic interactions between the two spheres is weak in this limit and each sphere behaves  approximately   as a single rotating sphere which does not propel. In Fig.~\ref{fig:separated}a, the variation of the propulsion speed as a function of the separated distance is calculated for different fixed values of  $r^*$. The propulsion speed decays as expected for large $h^*$. Interestingly, a non-monotonic variation occurs when the spheres are close to each other (small $h^*$). The swimming speed first increases with $h^*$, reaching a maximum around $h^* \approx 2.5$, before decaying to zero with further increase in $h^*$. In Sec.~\ref{sec:physical}, a  simple physical explanation to this  non-monotonicity is discussed; the dotted line (black online) in Fig.~\ref{fig:separated}a corresponds to the predictions by a simplified analytical model based on this  explanation.

For separated spheres, we can again vary the radii ratio, $r^*$, at different fixed separated distance $h^*$ (Fig.~\ref{fig:separated}b) and results similar to the case of touching spheres  is observed: for any value of $h^*$ there exists an optimal value of $r^*$ at which the dimensionless propulsion speed reaches a maximum. The simplified model (Sec.~\ref{sec:physical}) again captures this trend qualitatively (dotted line - black online, Fig.~\ref{fig:separated}b).

Finally, by plotting the isovalues of the propulsion speed as a function of both $r^*$ and $h^*$ (Fig.~\ref{fig:optimize}), we are able to optimize the snowman geometry for the overall maximum propulsion speed. The optimal geometry occurs at $(r^*, h^*) = (0.46, 2.5)$, and a schematic diagram of the optimal snowman is drawn to-scale in Fig.~\ref{fig:optimize}.

\begin{figure}[t]
\begin{center}
\includegraphics[width=0.45\textwidth]{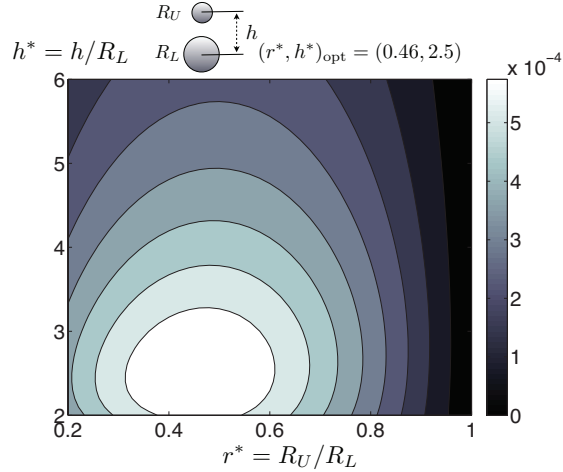}
\end{center}
\caption{\label{fig:optimize} Optimization snowman propulsion. Iso-values of the dimensionless propulsion speed with dimensionless separation distance, $h^*$, and ratio of sphere radii, $r^*$. The optimal geometry for maximum propulsion speed is given by $(r^*, h^*)_{\text{opt}} = (0.46, 2.5)$. A schematic diagram showing the optimal geometry is drawn to scale above.}
\end{figure}

\section{Microrheology via snowman}\label{sec:rheology}

\subsection{Scaling}

In the sections above we  have derived an analytical expression, valid in the small $\De$ regime, relating the propulsion speed to the intrinsic properties of the complex fluid, namely the normal stress coefficients (Eq.~\ref{eqn:finalRe}). Turning all dimensionless variables back in dimensional form, this relationship reads formally
\begin{align}
U  =   \left( C^\text{S}_1 \Psi_1 + C^\text{S}_2  \Psi_2 \right) \frac{R_L \Omega^2}{\eta}, 
\label{eqn:dimensionalU}
\end{align}
where $C_1^{\text{S}}$ and $C_2^{\text{S}}$ are dimensionless coefficients depending solely on the snowman geometry ($h^*$ and $r^*$) and defined  as
\begin{align}
C^{\text{S}}_1 &=  \frac{ \int_{V_f^*}  {\stackrel{\triangledown}{\dot{\boldsymbol{\gamma}_0^*}}}: \nablaB^* \u^*_{\text{aux}}dV^*}{2 F^*_{\text{aux}}}, \label{eqn:C1S}\\
C^{\text{S}}_2 &= - \frac{ \int_{V_f^*} \left(  \gammaB_0^* \cdot \gammaB_0^* \right) : \nablaB^* \u^*_{\text{aux}}dV^*}{F^*_{\text{aux}}}\cdot \label{eqn:C2S}
\end{align}

Since the second normal stress coefficient $\Psi_2$ is usually much smaller than the first normal stress coefficients $\Psi_1$, we might ignore $\Psi_2$ and obtain an estimation of $\Psi_1$ by measuring the propulsion speed of a snowman $U$, \textit{i.e.}
\begin{align}
\Psi_1 \approx \frac{U}{C_1^\text{S}} \frac{\eta}{R_L \Omega^2},
\end{align}
where $C_1^{\text{S}}$ depends only on geometry and can be computed using Eq.~\eqref{eqn:C1S}. This expression demonstrates the use of  locomotion ($U$) to probe the local non-Newtonian properties of the fluid ($\Psi_1$).

\subsection{Second Experiment: Repulsion of two equal spheres}\label{sec:equalSphere}

In scenarios where both values of $\Psi_1$ and $\Psi_2$ are desired, a second  experiment is necessary to obtain a second, independent, measurement of a combination of the normal stress coefficients. We propose to measure in the second experiment the relative speed (repulsion) of two rotating equal spheres of radius $R_E$, with their centers separated by a distance $h$ (see Fig.~\ref{fig:equalSphere} inset for notations and geometry). Should the two equal spheres be connected as a rigid body, no propulsion would occur by symmetry. However, if  the equal spheres  are not connected but allowed to freely translate along their separation axis, upon imposing rotation they will translate with velocities of equal magnitude but opposite directions provided the fluid is non-Newtonian.

We adopt the same non-dimensionalizations as previous sections (all lengths are now scaled by $R_E$) and drop the stars for simplicity; all variables in this section are dimensionless unless otherwise stated. Denoting the dimensionless velocity of the lower sphere as $\V$, we again expand the repulsion velocity in powers of $\Deso$, $\V = \Deso \V_1 + O(\Deso^2)$, and determine the first order velocity $\V_1$ using our use of the  reciprocal theorem as described in Sec.~\ref{sec:FirstOrder}. By symmetry, the upper sphere translates with velocity $-\V$ (equal speed but opposite direction as the lower sphere).

In this scenario we have to again define two setups, one for the main problem and one for the auxiliary problem. For the main problem, we consider the rotational motion of two free 
equal spheres about their line of  centers \cite{jeffery15}. Since the motion is force-free ($\F_1^{\alpha} = 0$ at each instant), Eq.~\eqref{eqn:finalRe} simplifies to  
\begin{align}
- \OmegaB^1_{\text{aux}} \cdot \T^1- \OmegaB^2_{\text{aux}} \cdot \T^2 + \U_1^1 \cdot \F^1_{\text{aux}}+ \U_2^1 \cdot \F^2_{\text{aux}} =  \int_{V_f} \left[ \left( \Ugamma_{0} + B \gammaB_0 \cdot \gammaB_0 \right) : \nablaB \u_{\text{aux}} \right] dV,
\end{align}
where $\OmegaB_1^{\alpha} = 0$ for the same reason as explained in Sec.~\ref{sec:FirstOrder}.

For the auxiliary problem, we consider the Newtonian translational motion ($\OmegaB_\text{aux}^\alpha = 0$) of two equal spheres moving towards each other at the same speed and hence force, $\F_\text{aux}^1 = -\F_\text{aux}^2 = \F^E_{\text{aux}} $. The exact solution to this problem was found by Brenner \cite{brenner61} in bi-spherical coordinates. We therefore have
\begin{align}
 \left( \U_1^1-\U_1^2 \right) \cdot \F^E_{\text{aux}}=  \int_{V_f} \left[ \left( \Ugamma_{0} + B \gammaB_0 \cdot \gammaB_0 \right) : \nablaB \u_{\text{aux}} \right] dV.
\end{align}
Note that the main problem here is a special case of that considered in Sec.~\ref{sec:FirstOrder}, but the auxiliary problem is completely different. We however still use the same symbols as in Sec.~\ref{sec:FirstOrder} for simplicity.

By symmetry, the two equal spheres propel with equal speed in opposite directions $\U_1^2 = - \U_1^1 = \V_1$ , hence
\begin{align}
-2 \V_1 \cdot \F^E_{\text{aux}} =  \int_{V_f} \left[ \left( \Ugamma_{0} + B \gammaB_0 \cdot \gammaB_0 \right) : \nablaB \u_{\text{aux}} \right] dV.
\end{align}
Since the repulsion velocity $\V_1$ (with magnitude $V_1$) and the force in the auxiliary problem $\F_{\text{aux}}^E$ (with magnitude $F_{\text{aux}}^E$) both act vertically, the equation above can be rewritten as
\begin{align}
V_1 &= -\frac{1}{2 F^E_{\text{aux}}}\int_{V_f} \left[ \left( \Ugamma_{0} + B \gammaB_0 \cdot \gammaB_0 \right) : \nablaB \u_{\text{aux}} \right] dV, \label{eqn:equalSphere}
\end{align}
where a positive value of  $V_1$ represents repulsion.

The only dimensionless parameter in this second experiment is the ratio of the separation distance to the radius of the spheres, which we write as $h^* = h/R_L$. Using Eq.~(\ref{eqn:equalSphere}) we 
 calculate the repulsion speed ($V_1 >0$) as a function of the dimensionless separation $h^*$ (solid line - blue online, Fig.~\ref{fig:equalSphere},   for $\De = 0.1$ and $\zeta = 0.5$), and the results are found to be in excellent agreement with the Oldroyd-B calculations (dots - red online, Fig.~\ref{fig:equalSphere}).

\begin{figure}[t]
\begin{center}
\includegraphics[width=0.45\textwidth]{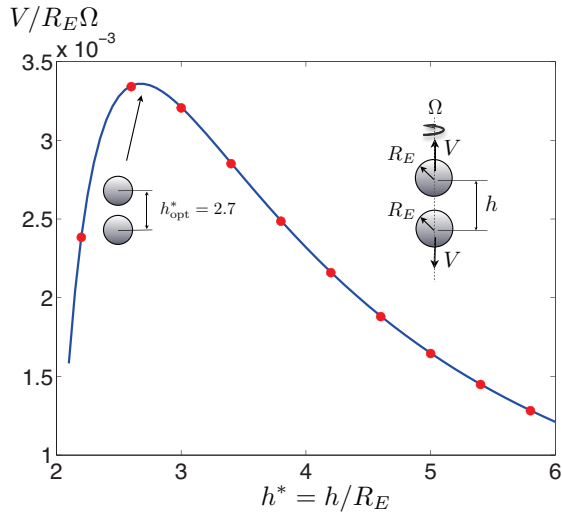}
\end{center}
\caption{\label{fig:equalSphere} Dimensionless repulsion speed, $V/R_E \Omega$, of two equal co-rotating spheres  as a function of their dimensionless separation distance, $h^*$, at $\De=0.1$ and $\zeta = 0.5$. Dots (red online): numerical results in an Odroyd-B fluid. Solid line (blue online): theoretical calculation for a second-order fluid.}
\end{figure}

Back to  dimensional variables, the leading order repulsion speed is formally given by
\begin{align}
V = \left( C_1^\text{E} \Psi_1 + C_2^{\text{E}} \Psi_2 \right) \frac{R_E \Omega^2}{ \eta}, \label{eqn:dimensionalV}
\end{align}
where $C_1^{\text{E}}$ and $C_2^{\text{E}}$ are dimensionless coefficients evaluated with the solution to the main and auxiliary problems described above
\begin{align}
C^{\text{E}}_1 &=  -\frac{ \int_{V_f^*}  {\stackrel{\triangledown}{\dot{\boldsymbol{\gamma}_0^*}}}: \nablaB^* \u^*_{\text{aux}}dV^*}{4 F^*_{\text{aux}}}, \label{eqn:C1E}\\
C^{\text{E}}_2 &= \frac{ \int_{V_f^*} \left(  \gammaB_0^* \cdot \gammaB_0^* \right) : \nablaB^* \u^*_{\text{aux}}dV^*}{2F^*_{\text{aux}}} \cdot \label{eqn:C2E}
\end{align}

\subsection{Determination of normal stress coefficients}

From measuring both the propulsion speed $U$ of a snowman (given by Eq.~\ref{eqn:dimensionalU}) and repulsion speed $V$ of the equal spheres (given by Eq.~\ref{eqn:dimensionalV}), we now have enough information to  deduce both the first and second normal stress coefficeints ($\Psi_1,\Psi_2$). If we choose the same radius for the lower sphere in both experiments $R_E = R_L$ (we use $R_L$ hereafter), we can write Eqs.~\eqref{eqn:dimensionalU} and \eqref{eqn:dimensionalV} in a matrix form as
\begin{align}
\begin{pmatrix}
U\\
V
\end{pmatrix}
=
\begin{pmatrix}
C_1^\text{S} & C_2^\text{S} \\
C_1^\text{E} & C_2^\text{E}
\end{pmatrix}
\begin{pmatrix}
\Psi_1 \\
\Psi_2
\end{pmatrix}
\frac{R_L \Omega^2}{\eta} \label{eqn:C},
\end{align}
where we denote by $\mathbf{C}$ the matrix containing the dimensionless coefficients ($C_1^\text{S},C_2^\text{S}, C_1^\text{E},C_2^\text{E}$) in Eq.~\eqref{eqn:C}.  The matrix $\mathbf{C}$ depends only on three geometric parameters, namely the ratio of the radii of the spheres in the snowman ($r^*=R_U/R_L$), the dimensionless separation distance in the snowman ($h^*_{\text{S}} = h/R_L$) and that for the equal spheres in the second experiment ($h^*_{\text{E}}=h/R_E$). The coefficients of the matrix can be readily computed via Eqs.~\eqref{eqn:C1S}--\eqref{eqn:C2S}  and Eqs.~\eqref{eqn:C1E}--\eqref{eqn:C2E}, and thus the matrix in Eq.~\eqref{eqn:C} can be inverted to obtain the values of $\Psi_1$ and $\Psi_2$.

For practical implementation of this microrheological technique, measurement errors in the velocity of the snowman are inevitable and depend on the specific equipment employed for tracking the motion of the probe. However, the geometry of the snowman can be designed so that $\Psi_1$ and $\Psi_2$ are insensitive to measurement errors in the velocities $U$ and $V$. The condition number (CN) of the matrix $\mathbf{C}$ to be inverted represents the maximum amplification factor of the relative measurement errors. The maximum relative errors in the normal stress coefficients would be equal to the condition number multiplied by the maximum relative measurement error. A small condition number is therefore desired. Similarly to the study by Khair and Squires \cite{khair}, we now investigate the value of condition number as a function of the geometry. To simplify the parametric studies, we first adopt the same separation distance in the first and second experiments ($h_\text{S}^* = h_\text{E}^*=h^* = h /R_L$), and explore the dependence of the condition number on $r^*$ and $h^*$, with results shown in Fig.~\ref{fig:CondNo}a. The condition number does not vary monotonically with the parameters, which implies that optimization is possible. Under this requirement and within the ranges of values considered ($r^* \in [0.2, 0.98]$ and  $h^* \in [2.1, 4]$), the geometry yielding the lowest condition number is $r^* = 0.46$ and $h^*=2.1$ (the corresponding condition number for $\mathbf{C}$ is $\approx 27.6$). When the requirement of $h_\text{S}^* = h_\text{E}^*$ is removed, by examining all combinations of the parametric values within the ranges  ($r^* \in [0.2, 0.98]$, $h^*_\text{S} \in [2.1, 4]$, and $h^*_\text{E} \in [2.1, 4]$), the minimum CN obtainable appears to be  $\approx 25.7$ with $r^* = 0.46$, $h^*_\text{S} = 2.6$, and $h^*_\text{E} = 2.1$.

\begin{figure}[t]
\begin{center}
\includegraphics[width=0.9\textwidth]{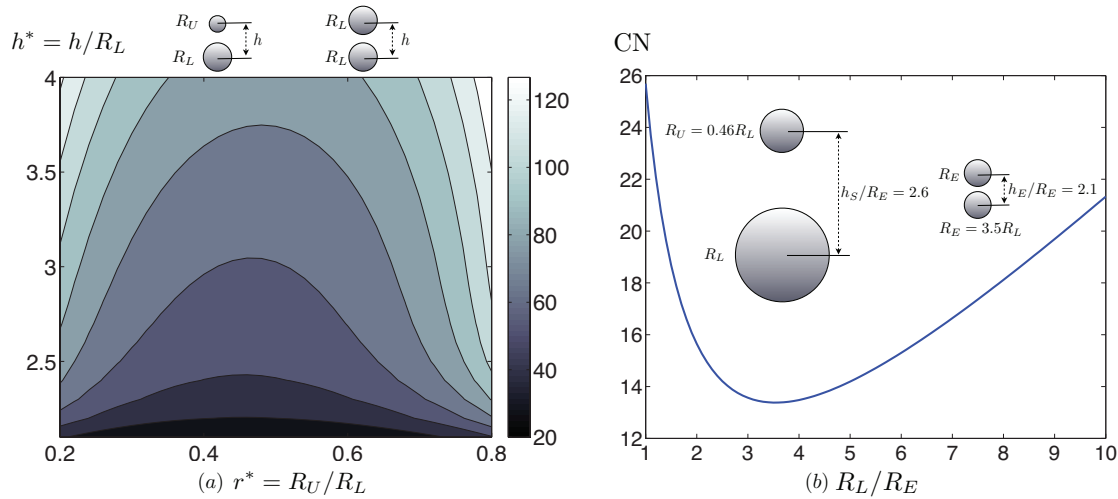}
\end{center}
\caption{\label{fig:CondNo} (a) Condition number (CN) for the matrix $\mathbf{C}$ as a function of sphere size ratio ($r^*$) and dimensionless separation distance ($h^*=h^*_\text{S}=h^*_\text{E}$). (b) CN as a function of $R_L/R_E$, for $r^* = 0.46$, $h^*_\text{S} = 2.6$, and $h^*_\text{E} = 2.1$.}
\end{figure}

The condition number can be further fine-tuned if we allow $R_L \neq R_E$, in which case we have the new matrix relation
\begin{align}
\begin{pmatrix}
U\\
V
\end{pmatrix}
=
\begin{pmatrix}
C_1^\text{S} & C_2^\text{S} \\
 C_1^\text{E}R_E/R_L & C_2^\text{E} R_E/R_L
\end{pmatrix}
\begin{pmatrix}
\Psi_1 \\
\Psi_2
\end{pmatrix}
\frac{R_L \Omega^2}{\eta} \cdot \label{eqn:Ctilde}
\end{align}
The modified dimensionless matrix $\tilde{\mathbf{C}}$ in Eq.~\eqref{eqn:Ctilde} now depends on one more parameter $R_L/R_E$, which is the ratio of the lower sphere radius in the snowman, $R_L$, to that of the equal sphere, $R_E$. In Fig.~\ref{fig:CondNo}b, we
investigate the dependence of the condition number of $\tilde{\mathbf{C}}$ with this new parameter, and adopt for all
other parameters the optimal geometric
 parameters we determined before ($r^* = 0.46$, $h^*_\text{S} = 2.6$, and $h^*_\text{E} = 2.1$). The variation
turns out to be also non-monotonic, and a minimum is achieved when $R_L/R_E = 3.5$ with $\text{CN} \approx 13.4$. A
schematic diagram  showing the corresponding geometrical setup of the two sets of experiment is given to scale in the inset
of Fig.~\ref{fig:CondNo}. The condition number could be brought further down with a full four-dimensional parametric
study and expanding the domains of the parametric studies. However, geometries yielding a lower $\text{CN}$ may
correspond to a negligible speeds undesirable for measurement. The current geometry ($r^* = 0.46$, $h^*_\text{S} = 2.6$,
$h^*_\text{E} = 2.1$, and $R_L/R_E = 3.5 $) has both a relatively low CN and a high propulsion speed, making it ideal
for experimental implementation. It is interesting to note that the optimal geometry for a small condition number we find here is close to
the optimal geometry producing the maximum propulsion speed for the snowman ($r^*=0.46$, $h_\text{S}^*=2.5$) determined
in Sec.~\ref{sec:separatedSphere}.

\section{Qualitative physical explanation}\label{sec:physical}

In this section, we turn to an explanation of the physical origin of the non-Newtonian propulsion of a snowman. Based on  physical intuition we present a simple model which successfully captures all the qualitative features of this mode of propulsion.

We first look into the simplest related problem, that of  a single sphere rotating in a complex fluid (a textbook problem discussed, for example in Ref.~\cite{bird1}). Non-Newtonian stresses lead to the creation of a secondary flow  in which the fluid moves towards the sphere in the equatorial plane and away from the sphere near the axis of rotation (see the inset of Fig.~\ref{fig:singleSphere} for an illustration of the secondary flow field) \cite{bird1}. 

This secondary flow can be understood  physically as a consequence of the hoop stresses  along the curved streamlines.  Polymer molecules in the fluid get stretched by the flow, leading to an extra tension along  streamlines. The presence of that extra tension along the closed circular streamlines leads to an inward radial contraction (like a stretched rubber band) pushing the fluid to thus go up vertically in both directions (by continuity) to produce the secondary flow. Notably, this secondary flow is independent of the direction of rotation of the sphere.

The argument for locomotion of the snowman is then the following. Based on the one-sphere result, we see that when the two spheres in a snowman are aligned vertically and subject to a rotation they  generate secondary flows and push against each other. For a single sphere, the strength of the secondary flow increases  with the size of the sphere \cite{bird1}. Consequently the smaller sphere is being pushed harder by the larger sphere than it is able to push against, and hence the two-sphere system is subject to a force imbalance, leading to propulsion. This physical understanding agrees with our results: propulsion always occurs in the direction of the smaller sphere, independently of the direction of rotation. Should the two spheres not be connected as a rigid body but free to translate vertically, they would repel each other, explaining physically our results in Sec.~\ref{sec:equalSphere}. 

Based on this intuitive argument, we can now construct a simple mathematical model. Using the same notations as above, for a sphere of radius $R$ rotating with an angular velocity $\Omega$ in a second-order fluid, the leading order solution $\v(\phi, r, \theta)$ in spherical coordinates  \cite{bird1} is $ \v^*\equiv \v/R \Omega = ( 1/r^* )^2 \sin \theta \  \e_\phi + \Deso (1-B) [ ( 1/2r^{*2} -3/2r^{*4} + 1/r^{*5}) (3 \cos^2 \theta -1) \ \e_r - 3 ( 1/r^{*4}-1/r^{*5} ) \sin \theta \cos \theta \ \e_\theta ]  + O \left(\Deso^2 \right)$. The Newtonian component of the flow field 
($\Deso = 0$) is the primary flow field, and it has only a azimuthal ($\phi$)  component. The secondary flow field, proportional to $\Deso$, is due to fluid elasticity and has only radial  ($r$) and polar ($\theta$) components.  As expected,  the dimensional secondary flow $\v$ is quadratic in $\Omega$, confirming our physical intuition that it should be  independent of the direction of rotation of the sphere. In the case where $B   < 1$ (recall that $B= -2 \Psi_2/ \Psi_1$), the relevant limit  for polymeric fluids,  the secondary flow occurs in the direction intuited  above and   shown  in the inset of Fig.~\ref{fig:singleSphere}. Note that the secondary flow field of a rotating single sphere would switch its direction when $B$ went above one, explaining  the switch in the propulsion direction of a snowman reported in Sec.~\ref{sec:FirstOrder} in that limit.

\begin{figure}[t]
\begin{center}
\includegraphics[width=0.5\textwidth]{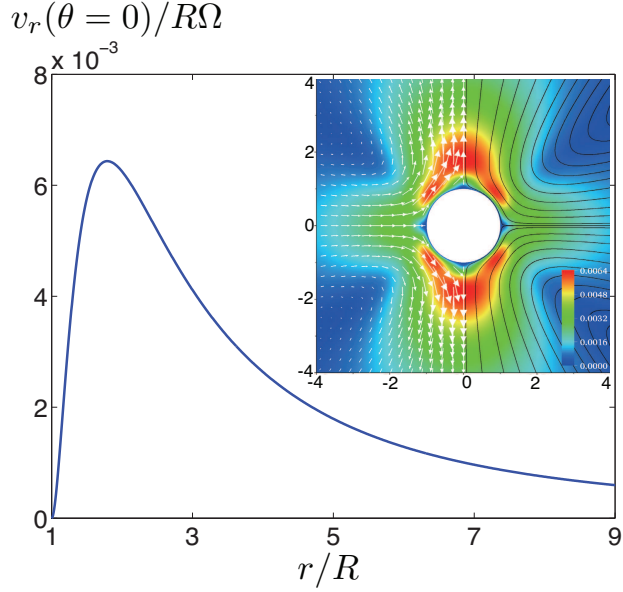}
\end{center}
\caption{\label{fig:singleSphere} Rotation of a single sphere in a second order fluid. Radial velocity along the rotation axis ($\theta = 0$) as a function of $r/R$ at $\De=0.1$ and $\zeta =0.5$. Inset: streamline pattern and velocity (shaded/color map) of the corresponding secondary flow.}
\end{figure} 

The dimensionless fluid velocity along the vertical axis ($\theta =0$) is given by $\v^*(r, \theta = 0) = \Deso \left( 1/r^{*2} -3/r^{*4} + 2/ r^{*5}    \right)  \ \e_r (\theta = 0)$,
where we have set $B =0$ to allow comparison with the numerical results.  The velocity along the vertical axis, shown in 
Fig.~\ref{fig:singleSphere},  is expected to display non-monotonic variation with the distance from the sphere  since the velocity decreases to  zero  both in the far field and on the solid surface. This is at the origin of the non-monotonic dependence of the snowman propulsion speed with the separation distance between the spheres shown in Fig.~\ref{fig:separated}a.

The forces experienced by the upper and lower spheres can be approximately estimated by considering the individual flow fields generated by their own rotation without the presence of the other sphere. We place an upper sphere at a distance $h^*=h/R_L$ from the center of the lower sphere. Using the same notations as in the previous sections, the dimensionless velocity generated by the lower sphere,  and evaluated at the location of the upper sphere, is given by $\v^{*}_L (h^*) = \Deso \left( 1/h^{*2} -3/h^{*4} + 2 h^{*5}    \right)  \ \e^L_r(\theta^L = 0)$, where $\e^L_r (\theta^L =0)$ is the unit radial vector in the polar direction $\theta^L = 0$, with respect to the coordinates system at the center of the lower sphere. Similarly, the dimensionless velocity generated by the upper sphere at the same distance,  $h^* $, but measured from the  center of the 
upper sphere is given by $\v^{*}_U (h^*) = r^* \Deso \left[ \left(r^*/h^*\right)^2 -3 \left( r^*/h^*\right)^4 + 2\left( r^*/h^*\right)^5    \right]  \ \e_r^U(\theta^U = \pi)$, where $\e^U_r (\theta^U =\pi)$ is the unit radial vector in the polar direction $\theta^U = \pi$, with respect to the coordinates system at the center of the upper sphere. Note that $\e^U_r (\theta^U =\pi) = -\e^L_r (\theta^L =0)$.
As a simple  approximation, we estimate the viscous drag force experienced by the upper and lower spheres to be $\mathbf{F}_U^* \sim  6\pi r^* \v_L^*(h^*)$ and $\mathbf{F}_L^* \sim  6\pi \v_U^*(h^*)$ 
respectively. The difference between  these two forces results in a net propulsive thrust. When dividing by an approximation of the translational resistance of the snowman at zero Deborah number with no  hydrodynamic interactions, $6\pi (1+r^*)$,  we obtain a simple estimate  of the dimensionless propulsion speed as $U^* \approx{|\mathbf{F}_U^*+\mathbf{F}_L^*|}/{6\pi(1+r^*)} $. This leads to
\begin{align}
U^* \approx  \frac{|r^* \v_L^*(h^*)+\v_U^*(h^*)|}{(1+r^*)} = \Deso \frac{r^* \left[ 3h^* (r^{*4}-1)-h^{*3}(r^{*2}-1)-2r^{*5}+2 \right]}{h^{*5} (1+r^*)} \cdot \label{eqn:simple}
\end{align}
In Eq.~\eqref{eqn:simple}, we verify that $U^*$ vanishes when $r^*=0$ (single sphere) and $r^*=1$ (equal spheres).  For the case of touching spheres ($h^* = 1+ r^*$), Eq.~\eqref{eqn:simple} simplifies to
\begin{align}
U^*_{\text{touch}} \approx \Deso \frac{2r^{*3}(1-r^*)}{(1+r^*)^6}  \cdot \label{eqn:simpleTouching}
\end{align}

Does this simple model capture the essential propulsion characteristics? In Fig.~\ref{fig:touchingR}, we plot the dimensionless  propulsion speed of a touching snowman estimated by this simple model (Eq.~\ref{eqn:simpleTouching}) as a function of $r^*$ (dotted line - black online) and compare with the theoretical  results from the reciprocal theorem approach  (solid line - blue online) and the numerical computations (symbols - red online). The simple model correctly predicts the order of magnitude and captures qualitatively the variation with $r^*$. For non-touching snowman, the qualitative model  (Eq.~\ref{eqn:simple}) also captures qualitatively the variation of the dimensionless propulsion speed with  $h^*$ (dotted line - black online - for $r^*=0.6$, Fig.~\ref{fig:separated}a), also predicting an optimal separation distance and therefore   supporting our understanding of a non-monotonic dependence  with $h^*$ as arising from the non-monotonicity of the single-sphere velocity (Fig.~\ref{fig:singleSphere}). As expected, Eq.~\eqref{eqn:simple} also captures the non-monotonic variation with respect to $r^*$ for separated snowman (dotted line - black online - for $h^* = 8$, Fig.~\ref{fig:separated}b).

\section{Discussion and Conclusions}\label{sec:discussion}

In this work, we present the design and mathematical modeling for a new non-Newtonian swimmer -- the snowman -- which propels only in complex fluids by exploiting asymmetry and the presence of normal stress differences under rotational actuation. The simple shape of our swimmer makes it ideally suited for experimental measurements. Note that if kept in place, the snowman would then act as a micro-pump for complex fluids.

The propulsion characteristics of the snowman are investigated by a combination of numerical computations (moderate values of $\De$ in an Odroyd-B fluid) and analytical treatment (small $\De$ in a second-order fluid). The underlying physics of propulsion, relying on  elastic  hoop stresses and geometrical asymmetry, is explained and based on this physical understanding a simple analytical model capturing all qualitative features  is successfully constructed. Note that since, as a  rule of thumb, inertial and elastic effects tend to produce secondary flows in opposite directions \cite{bird1}, we expect that an inertial (instead of viscoelastic) snowman should  swim in the opposite direction (from small to large sphere).

The two-sphere setup proposed in this work is arguably the simplest geometry able to swim in a complex fluid under uniform rotation. It of course  simplifies the analysis since the required Newtonian solutions to be used in our integral approach  are all available.  Any axisymmetric but top-down asymmetric geometry should also work, for example a cone, and clearly there remains room for shape optimization in that regard. Additionally, studying the snowman dynamics under a time-varying  rotation could lead to a rich dynamics with potentially non-trivial stress relaxation effects.

One of the  main ideas put forward in this work is the use of locomotion as a proxy to probe the local non-Newtonian properties of the fluid. The snowman can be used as a micro-rheometer to estimate the first normal stress coefficient on its own, or to measure both the first and second normal stress coefficients with the help of another complementary experiment.  Khair and Squires \cite{khair} recently proposed to measure normal stress coefficients by pulling microrheological probes and measuring the relative forces on the probes. In our work, we propose alternatively to perform only kinematic measurements of the sphere speeds instead of forces, which could present an interesting alternative from an experimental standpoint.

We finally comment on  a potential experimental implementation of the snowman technique. We are aware of a number of rotational micro-manipulation techniques (see a short review in Ref.~\cite{bishop}). For example, spinning micro-particles may be achieved by the use of optical tweezers and birefringent objects \cite{friese}. Birefringence allows the transfer of angular momentum from the circularly polarized laser to the particle, producing controlled rotation. By rotating spherical birefringent crystals (vaterite), this technology has been implemented as a micro-viscometer to probe fluid viscosity \cite{bishop, knoner, parkin}.  A similar mechanism may be useful for the two-sphere setup in this work although simultaneous rotation of two spheres may introduce experimental challenges. Our dual-purpose snowman, both a micro-propeller and a micro-rheometer, invites experimental implementation and verification.

\section*{Acknowledgments}
Funding by the National Science Foundation (Grant No. CBET-0746285 to E.~L.) and the Croucher Foundation (through a scholarship to O.~S.~P.) is gratefully acknowledged.


\begin{thebibliography}{78}
\bibitem [{fno()}]{fnote}%
  \BibitemOpen
  \href@noop {} {}\bibinfo {note} {If the drag on the slender rod is taken into
  account in modeling, the overall force balance should include the small
  contribution from the drag on the rod, which will slightly decrease the
  propulsion speed of the snowman.}\BibitemShut {Stop}%
\end{thebibliography}
%

\end{document}